# Observation of non-Abelian nodal links in photonics

Erchan Yang[1,2], Biao Yang[3*], Oubo You[1,2], Hsun-chi Chan[1,2], Peng Mao[2], Qinghua Guo[4], Shaojie Ma[1,2], Lingbo Xia[2], Dianyuan Fan[1], Yuanjiang Xiang[5†], Shuang Zhang[2‡]

[1]International Collaborative Laboratory of 2D Materials for Optoelectronic Science & Technology of Ministry of Education, Institute of Microscale Optoelectronics (IMO), Shenzhen University, Shenzhen, 518060, China

[2]School of Physics and Astronomy, University of Birmingham, Birmingham B15 2TT, United Kingdom

[3]College of Advanced Interdisciplinary Studies, National University of Defense Technology, Changsha 410073, China

[4]Department of Physics and Institute for Advanced Study, The Hong Kong University of Science and Technology, Hong Kong, China

[5]School of Physics and Electronics, Hunan University, Changsha 410082, China

*Corresponding author.

yangbiaocam@nudt.edu.cn (B.Y.)

†Corresponding author.

xiangyuanjiang@126.com (Y.X.)

‡Corresponding author.

s.zhang@bham.ac.uk (S.Z.)

**Abstract**

In crystals, two bands may cross each other and form degeneracies along a closed loop in the three-dimensional momentum space, which is called nodal line. Nodal line degeneracy can be designed to exhibit various configurations such as nodal rings, chains, links and knots. Very recently, non-Abelian band topology was proposed in nodal link systems, where the nodal lines formed by consecutive pairs of bands exhibit interesting braiding structures and the underlying topological charges are described by quaternions. Here, we experimentally demonstrate non-Abelian nodal links in a biaxial hyperbolic metamaterial. The linked nodal lines threading through each other are formed by the crossings between three adjacent bands. Based on the non-Abelian charges, we further analyze various admissible nodal link configurations for the three-band system. On the interface between the metamaterial and air, surface bound states in the continuum (BICs) are observed, which serves as the symmetry-enforced derivative of drumhead surface states from the linked nodal lines. Our work serves as a direct observation of the global topological structures of nodal links, and provides a platform for studying non-Abelian topological charge in the momentum space.

The research on topological gapless phases focuses on investigation of various band degeneracies, such as Weyl points, Dirac points and nodal lines [1-16]. Nodal line semimetals have degeneracy along lines forming closed loop(s) in the three-dimensional momentum space [2,9]. Their existence usually requires the presence of certain symmetries, such as space-time inversion (*PT*), chiral and mirror symmetries [9,17-19]. When certain symmetries of the system are broken, nodal line semimetals can transform into other topological phases, such as Weyl, Dirac semimetals and 3D topological insulators [3,20-23]. In the last few years, the concept of nodal lines has also been extended to classical systems, leading to fruitful research in photonics and phononics/acoustics [14,19,24-26].

New interests are inspired when considering global linking or knotting properties of nodal lines [17,27-35]. Nodal lines have been classified into topological nodal rings, nodal chains [8,14,36-38], nodal links [17,30-34] and nodal knots [28,29]. Nodal rings, or the ordinary nodal loops, are relatively trivial due to the lack of global topological structures. A nontrivial representative of nodal lines is nodal link, namely, several nodal lines nontrivially linked with each other. In comparison, a nodal knot represents a single nodal line 'linked' with itself [28,29]. With multiple nodal lines touching together, nodal chains serve as the transition phase between nodal loops and nodal links/knots [35]. Until now, all of the demonstrated nodal line structures belong to $\mathbb{Z}$ or $\mathbb{Z}_2$ classes, the Abelian homotopy groups. They are well known in topological physics, manifested as the Chern numbers or the winding numbers. Very recently, researchers [39] proposed non-Abelian topological charges of nodal links in non-interacting metals, which exhibit braiding topological strucutres and trajectory dependent node collisions.

Here, we experimentally demonstrate non-Abelian topological charges in biaxial hyperbolic metamaterials, which host linked nodal lines formed by three adjacent bands. The non-Abelian topological charges underly various admissible nodal line configurations and the corresponding topological phase transitions between them. Our research unveils the connection between the linearly polarized eigenstates of the system and non-Abelian global band topology. Interestingly, for the three-band $PT$ photonic system studied here, the three linearly polarized eigenstates (one longitudinal and two transverse) at each point in the momentum space can be mapped to the three axis of a biaxial nematics ellipsoid molecule, while the nodal lines with $\pi$ disclination closely resemble the nematics line defect in three-dimensional real space [39,40].

We further observe surface bound states in the continuum (BICs) [41-45] in our linked nodal line system. BICs are isolated and spatially bounded states with frequencies locating inside the continuum and coexist with extended waves. They have attracted great interests recent years in optical systems because they are deemed to hold great promise in applications including lasers, sensing, fluorescence enhancement and light-matter interactions, benefitting from the long lifetime and strong localization [44,46].

We start from an effective medium description of the biaxial hyperbolic metamaterial. The effective relative permittivity tensor takes the form of $\overleftrightarrow{\varepsilon} = diag[\varepsilon_x(\omega), \varepsilon_y(\omega), \varepsilon_z(\omega)]$, where along each direction the system exhibits a Drude type of dispersion as $\varepsilon_i(\omega) = \varepsilon_{i\infty}\left(1 - \frac{\omega_{pi}^2}{\omega^2}\right)$, where $i = x, y, z$ indicates the three orthogonal directions. Along all three directions the relative permittivities $\varepsilon_{x,y,z}$ are mutually different to realize biaxial electromagnetic responses. The system can be described by an effective Hamiltonian taking the form as,

$$\mathcal{H}(k_x, k_y, k_z)\vec{\psi} = \omega^2 \vec{\psi} \quad (1)$$

where,

$$\mathcal{H} = \begin{pmatrix} \frac{k_y^2+k_z^2}{\varepsilon_{x\infty}} + \omega_{px}^2 & -\frac{k_x k_y}{\sqrt{\varepsilon_{x\infty}\varepsilon_{y\infty}}} & -\frac{k_x k_z}{\sqrt{\varepsilon_{x\infty}\varepsilon_{z\infty}}} \\ -\frac{k_x k_y}{\sqrt{\varepsilon_{x\infty}\varepsilon_{y\infty}}} & \frac{k_x^2+k_z^2}{\varepsilon_{y\infty}} + \omega_{py}^2 & -\frac{k_y k_z}{\sqrt{\varepsilon_{y\infty}\varepsilon_{z\infty}}} \\ -\frac{k_x k_z}{\sqrt{\varepsilon_{x\infty}\varepsilon_{z\infty}}} & -\frac{k_y k_z}{\sqrt{\varepsilon_{y\infty}\varepsilon_{z\infty}}} & \frac{k_x^2+k_y^2}{\varepsilon_{z\infty}} + \omega_{pz}^2 \end{pmatrix} \quad (2)$$

and $\vec{\psi} = (\sqrt{\varepsilon_{x\infty}}E_x, \sqrt{\varepsilon_{y\infty}}E_y, \sqrt{\varepsilon_{z\infty}}E_z)$. The real diagonal form of $\overleftrightarrow{\varepsilon}$ indicates that the model possesses time-reveral symmetry $T$, three orthogonal mirror symmetries $M_x$, $M_y$, $M_z$, and thus inversion symmetry $P$. Under the current basis, these mirror symmetries are represented by $M_x = [-1,1,1]$, $M_y = [1,-1,1]$ and $M_z = [1,1,-1]$. $PT = K$ with $K$ being complex conjugate indicates $\mathcal{H}^*(\vec{k}) = \mathcal{H}(\vec{k})$, ensuring that the eigen electromagnetic states are all linearly polarized.

For simplicity, $\omega_{px}$ is normalized to 1 in the following discussion. First we set $\omega_{py,z} = 0$ to further simplify the system for the study of the non-Abelian topological charges. Figure 1(a) shows the nodal links formed by three adjacent bands, where the crossing between lower(upper) pair of bands is coloured in red(blue). Due to mirror symmetry, these nodal lines are all located on the high-symmetry planes (centre planes, $k_i = 0$ with $i = x, y, z$). On the plane $k_y = 0$, the degeneracy between the lower two bands continuously maps a $\infty$-shaped nodal line (red). While the nodal rings (blue) formed by upper two bands threading through each hole of $\infty$-shaped nodal line are located on the $k_z = 0$ plane. Thus, two sets of nodal lines mutually link each other. In addition, there are two nearly straight nodal lines (blue) located on the $k_x = 0$ plane, which tangentially touch the blue nodal rings, forming two nodal chains (cyan dots in Fig. 1(a)) in the $k$-space. It will be clear that the two straight nodal lines are non-Abelian charges induced. (see Suppl. Section 1 and Fig. S1 for the nodal link band structure [47])

In Fig. 1b, we present the three-band dispersion along $k_x$ axis and the corresponding degeneracy points. The dispersion of the one-dimensional band structure features a flat band of longitudinal mode (L) with $\omega_L = 1$ and two transverse modes, $T_1$ and $T_2$, with linearly polarizing along $z$ and $y$ directions, respectively. The mirror eigenvalues of the longitudinal mode and the two transverse modes $T_1$ and $T_2$ are given by (+, +), (+, -) and (-, +) respectively, where the first and the second indices represent the eigenvalues of $M_y$ and $M_z$, respectively [Fig. 1(b)]. Here, it is worth mentioning that the second band behaves as a common band (thick) in forming the linked nodal lines with the other two.

The global nodal line configurations can be characterized by non-Abelian topological charges [39]. (see Suppl. Section 2 for the Abelian charge analysis of the system [47]) For any loops (such as green loops in Fig. 1(a)) in the momentum space that do not touch any nodal lines, the Hamiltonian can be flattened as $Q(\vec{k}) = W(\vec{k}) I_{1,1,1} W^T(\vec{k})$, where $I_{1,1,1} = diag[1,2,3]$. The parameter space of the Hamiltonian is isomorphic to the coset space $M_3 = \frac{O(3)}{O(1)^3}$, where $O(N)$ is the orthogonal group and $O(1) \equiv \pm 1$, meaning that multiplying any of the eigenstates with $\pm 1$ does not change the form of the Hamiltonian. Its fundamental homotopy group is non-Abelian quanternion group $\pi_1(M_3) = \mathbb{Q}$, where $\mathbb{Q} = [\pm 1, \pm i, \pm j, \pm k]$ with anticommuting imaginary units satisfying $i \cdot j = -j \cdot i = k$ and $i^2 = j^2 = k^2 = -1$ [39,40]. In the three-band system, $\pm i$ ($\pm k$) characterize closed loops that encircle a blue (red) nodal line formed between the upper (lower) two bands (Fig. 1(c) and (d)), whereas $\pm j$ corresponds to the loops enclosing both a blue and a red nodal line (Fig. S2(g)). The sign of the charge assigns an orientation to the nodal lines (Fig. 2). For a loop encircling two nodal lines (blue) of the same orientation between the same pair of bands, the topological charge is "$-1$", which is topologically distinct to the trivial class "1", which can be shrunk to a single point without touching any nodal lines. (More analysis of the non-Abelian topological charges is given in Suppl. Section 2 [47]).

These non-Abelian topological charges further impose several constraints on admissible nodal line transitions. We start from a relatively simple case [Fig. 2(a)] and predict the global evolution of nodal links [Fig. 2(b-d)]. Fig. 2(a) shows a simple configuration consisting of a ∞-shaped nodal line (red) formed between the first and second bands with its two loops respectively threaded by two nodal lines formed between the second and third bands (blue), wherein the orientation of a nodal line is reversed each time it goes under a nodal line of the other color due to the non-trivial braiding rules arising from the noncommutativity of the quaternion charge [39]. By adjusting the material parameters, the two blue nodal lines can join each other and form a loop, meanwhile creating two nearly straight nodal lines on the $k_x = 0$ plane [Fig. 2(b)] to avoid orientation (indicated by arrows) singularities at the chain points (cyan dots in Fig. 2(b)). Thus, the emergence of the additional nodal lines is protected by the underlying global band topology. Figure 2(c) shows another non-trivial case, wherein a small red nodal ring is generated in the middle, forming two chain points (cyan points protected by both $M_y$ and $M_z$) with the other two circular rings. (see Suppl. Section 3 and Fig. S4 [47]). The configuration in Fig. 2(a) can also be transformed into that shown in Fig. 2(d), where the chain point at the origin disappears and the two touched nodal rings merged into a single ring. This can be explained by considering the green trivial loop lying in $k_x$-$k_y$ plane shown in Fig. 2(d). For the same loop in Fig. 2(a), the red nodal lines with opposite orientations threading through it would cancel out the topological charges, resulting a trivial winding of the loop, which is consistent with Fig. 2(d). The above analysis further confirms the underlying global non-abelian topology of the nodal lines.

We experimentally design a microwave metamaterials to demonstrate the nodal links as shown in Fig. 1(a). It consists of an array of thin metallic wires and metallic cross structures patterned on a dielectric plate, as shown schematically in Fig. 3(a) (space group No. 47: Pmmm). The relative permittivity of the dielectric plate is 4.1. The continuous metallic wire provides the Drude-like dispersion along $x$ direction. The cross structures with arms along $x = \pm y$ orientations provide the anisotropy in the $y$ and $z$ directions. The entire sample contains $75 \times 75 \times 52$ unite cells with periods $p_x = p_y = 4mm$ and $p_z = 3mm$. The bulk modes propagating in the metamaterial is simulated by using a commercially available software - CST Microwave Studio. The simulated dispersion curve along $k_x$ direction is shown in Fig. 3(b). The dispersion appears similar to that of the effective medium in Fig. 1(b), except for the presence of dispersion in the longitudinal mode due to nonlocal effect [48]. The locations of nodal lines in the $k$-space obtained via simulation are shown in Fig. 3(c) (solid lines). Being the same to the effective media analysis (Fig. 1(a)), the realistic metamaterial structure possesses two nodal lines that are linked to each other.

The bulk states of the metamaterial are probed by using the near-field scanning setup ([47], Section 7, Fig. S8). The measured complex fields are subsequently Fourier transformed to obtain their distribution in the $k$-space. The projected Equi-Frequency Contours (EFCs) onto the $k_x$-$k_z$ plane at three different frequencies are shown in Fig. 3(d)-3(e). As a reference, the projection of the boundary of the bulk states simulated by CST are also provided in the form of white solid lines. At each frequency, four nodal points together with the eye-shaped voids are clearly visible in the EFC. The continuous variation of these nodal points over frequency forms the ∞-shaped nodal line on the $k_y = 0$ plane in Fig. 3(c). The projected EFCs onto the $k_x$-$k_y$ plane at three higher frequencies are shown in Fig. 3(f)-3(g), which features two elliptically shaped contours intercepting each other at four points. The continuous variation of the four interception points at different frequencies form the circular nodal line on the $k_z = 0$ plane in Fig. 3(c). Similarly Fig. 3(h) and (i) show the EFCs on $k_z$-$k_y$ plane. More experimental EFCs results further confirming the presence of linked nodal lines are shown in Fig. S9-S11 ([47], Section 8). Their positions are labelled as squares in Fig. 3(b). Hence the existence of the nodal links is verified experimentally.

For systems possessing nodal line degeneracies, there may exist drumhead surface states [2,9,12,49,50]. To explore the surface states of the metamaterial, a slightly different measurement configuration is employed [[47], Section 7, Fig. S8(d) and S8(e)]. We measure the surface state for both the interfaces in the $x$-$z$ plane and that in the $x$-$y$ plane. For the interface in the $x$-$z$ plane, the dispersion of surface state along $k_x$ direction together with the simulation result is shown in Fig. 4(a) (see the analytical results based on the effective media analysis in [47], Section 4). It is shown that the dispersion of the surface state along the $k_x$ direction is asymptotical to the free space dispersion of air at very low frequencies. It becomes more curved at higher frequencies and eventually terminates at point $P_2$. The dispersion curve is divided into two sections, that with $k_x$ less than the $T_1$ mode (coloured in green), and that with $k_x$ greater than the $T_1$ mode (coloured in magenta). These two sections in green and magenta colour represent the surface BIC and normal surface state, respectively. Both the simulated and measured dispersion curves show similar features as that calculated based on effective medium approximation ([47], Section 4-5, Fig. S6(a)). The probed EFCs of the surface states with different frequencies are shown in Fig. 4(c)-4(e). At each frequency, there are two bright spots that correspond to the surface BICs, whose locations coincide well with the simulation results, as indicated by the green circular points. With the increase of frequency, the surface BICs move further away from the light cone, indicating a stronger confinement on the air side. Detailed analysis of the surface states in [47] (Section 5) shows that the surface BICs are essentially the symmetry-enforced extension of the drumhead surface states.

For the interface in the $x$-$y$ plane, the measured and simulated dispersion curves of the surface BIC along $k_x$ direction are shown in Fig. 4(b). The surface BIC from both the simulation and the measurement connects between $\Gamma$ point and $P_1$ point, which agrees well with that of the effective medium ([47], Section 5, Fig. S6(e)). Figure 4(f)-4(h) present the surface states at three different frequencies. The surface BICs are indicated by the isolated bright spots on the $k_x$ axis. Besides the surface BIC, the normal surface state in the form of Fermi arcs are also observed that connect between the projections of the nodal points and the EFC of the bulk states. Again, the above features agree very well with the effective medium description provided in [47] (Section 5, Fig. S6). Detailed analysis about propagating constants of surface states is shown in [47] (Section 6, Fig. S7).

Compared with previously demonstrated nodal lines [25,51], the nodal link formed by three bands exhibits rich global topological structures. It provides a controllable platform towards investigating various nodal line phase transitions with interesting non-abelian global topology, such as quaternion charges [39,52]. In addition, it also introduces new interesting phenomena that arise from the unique optical density of states of the linked structure, such as spontaneous emission, resonant scattering, black body radiation[53]. Furthermore, due to the $\pi$ disclination of the polarization state around each nodal point, the reflection or transmission field is expected to carry momentum space vortex[54]. By introducing synthetic vector potential, intriguing quantum oscillation in the nodal links may also be observed[34]. The observed BIC states may facilitate applications including lasing, sensing and surface beam shaping and wavevector selection[44].

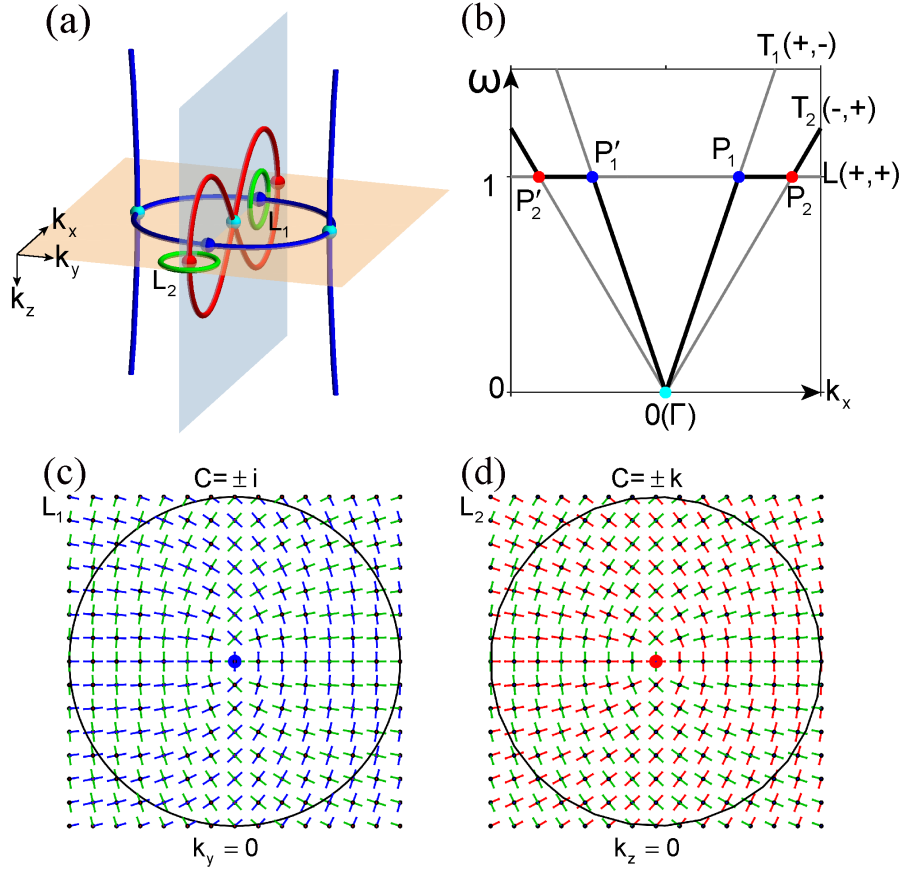

FIG. 1. Nodal links in biaxial hyperbolic metamaterials. (a) Nodal links in biaxial hyperbolic metamaterials. The effective parameters of the system are $\varepsilon_{x\infty} = 11/2$, $\varepsilon_{y\infty} = 6$, $\varepsilon_{z\infty} = 2$, $\omega_{px} = 1$ and $\omega_{p,z} = 0$. The red/blue nodal lines are formed by the lower/upper pair of bands. Green circles indicate the closed loops enclosing the nodal lines. (b) Band structure along $k_x$ direction shows four crossing points ($P_1'$, $P_2'$, $P_1$, $P_2$) between one longitudinal mode L and two transverse modes $T_1$ and $T_2$. The second band is highlighted by thick black line. $\omega$ is normalized by $\omega_{px}$. (c) Calculated linear-polarized electric field frame on the $k_y = 0$ plane around the loop $L_1$. (d) Similar to (c) but on the $k_z = 0$ plane around the loop $L_2$. The red/green/blue corresponds to 1st/2nd/3rd band, which also defines the right-handed frame. The corresponding non-Abelian topological charges are labelled.

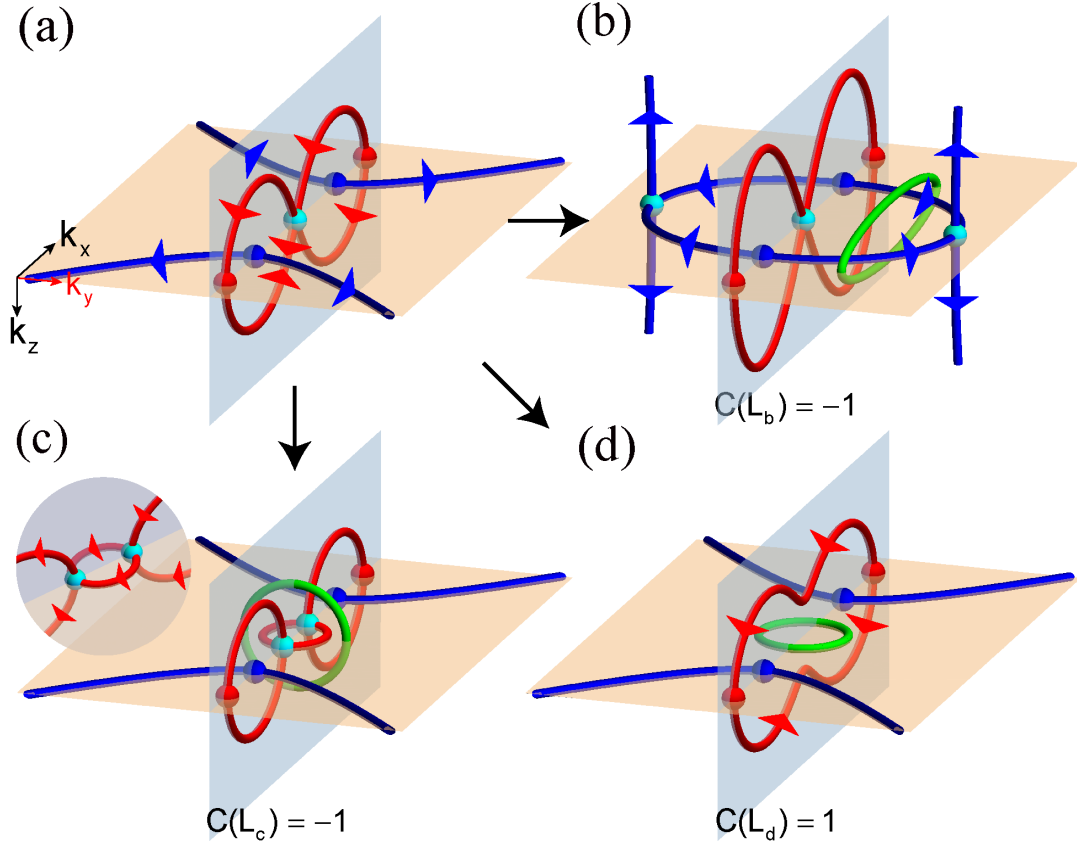

FIG. 2. Admissible nodal line configruaions protected by non-Abelian topological charges. The corresponding parameters are: (a)$\varepsilon_{x\infty} = 1$, $\varepsilon_{y\infty} = 6$, $\varepsilon_{z\infty} = 2$, $\omega_{px} = 1$ and $\omega_{py,z} = 0$; (b) $\varepsilon_{x\infty} = 11/2$, $\varepsilon_{y\infty} = 6$, $\varepsilon_{z\infty} = 2$, $\omega_{px} = 1$ and $\omega_{py,z} = 0$; (c) $\varepsilon_{x\infty} = 1$, $\varepsilon_{y\infty} = 6$, $\varepsilon_{z\infty} = 2$, $\omega_{px} = 1$, $\omega_{py} = 1/4$ and $\omega_{pz} = 0$; (d) $\varepsilon_{x\infty} = 1$, $\varepsilon_{y\infty} = 6$, $\varepsilon_{z\infty} = 2$, $\omega_{px} = 1$, $\omega_{py} = 0$ and $\omega_{pz} = 1/6$; The orientations are indicated in (a) by arrows. The orientation of a nodal line is reversed each time it goes under a nodal line of the other color. (b-d) show the admissible nodal line transitions. The corresponding non-Abelian topological charges and oreintations (arrows) are labelled. (More details in Fig. S4 [47])

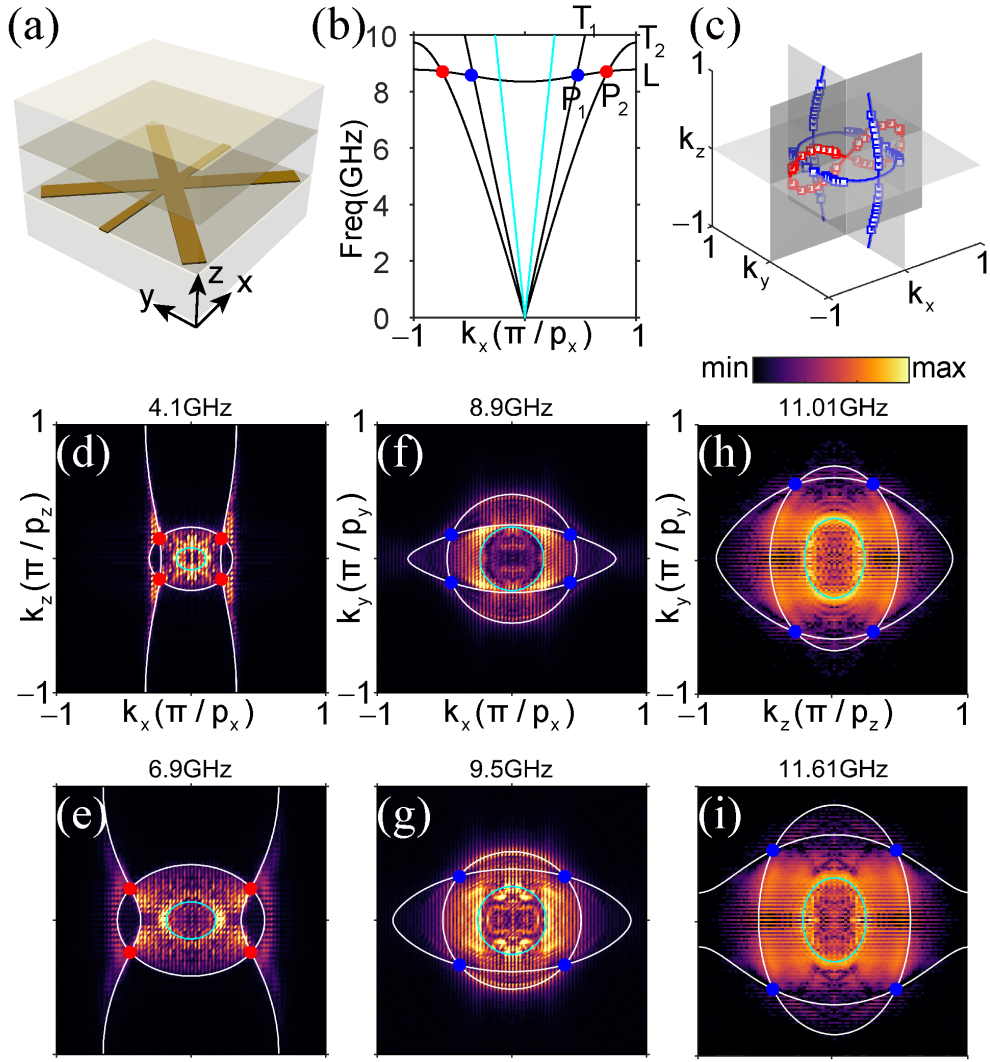

FIG. 3. Experimental demonstration of the nodal links in biaxial hyperbolic metamaterial. (a) Schematic diagram of the unit cell, belong to the space group No. 47: Pmmm. The metallic structure is made of copper with a thickness of 0.035mm. The width of the connective metallic wire is 0.2mm. The length and width of each arm of the metallic cross structure are 5mm and 0.4mm, respectively. (b) The simulated bulk dispersion curve along $k_x$ direction. (c) The experimentally probed (square dots) and the numerically simulated nodal lines (solid lines) in the $k$-space. (d)-(e), (f)-(g) and (h)-(i) The projected Equi-Frequency Contours (EFCs) with respect of different frequencies on the $k_x$-$k_z$, $k_x$-$k_y$ and $k_z$-$k_y$ planes, respectively. White/cyan solid line corresponds to the projected bulk state boundary of the biaxial hyperbolic metamaterial/air.

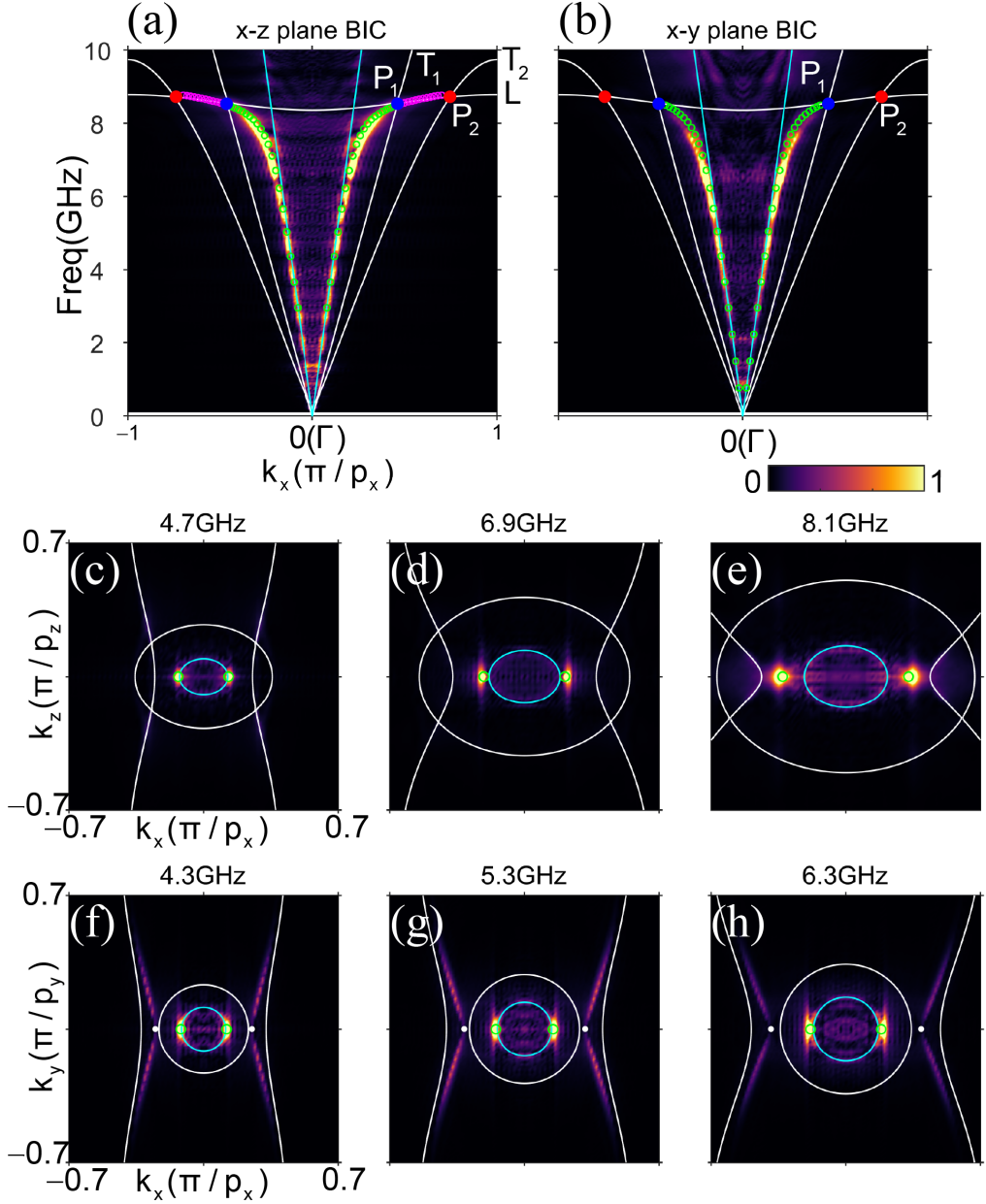

FIG. 4. Experimental observation of surface bound state in the continuum (BIC). (a), (b) The measured dispersion of surface state along $k_x$ direction in the interfaces of $x$-$z$ and $x$-$y$ planes, respectively. Green and magenta circles indicate the simulated surface BICs and normal surface states, respectively. (c)-(e), (f)-(h) Equi-Frequency Contours (EFCs) of surface states with respect of three different frequencies on the interfaces $x$-$z$ and $x$-$y$ planes, respectively. The white dots in (f)-(h) denote the projection of nodal points. Green circles indicate the surface BICs. White/cyan solid line corresponds to the projected bulk state boundary of the biaxial hyperbolic metamaterial/air.

18